\documentclass[aps,pra,reprint,amsmath,amssymb,superscriptaddress,nobibnotes, longbibliography]{revtex4-1}
\usepackage{graphicx,SIunits}
\usepackage{bm}     % bold math
\usepackage{hyperref} % add hypertext capabitlities
\usepackage{dsfont}    % allows \mathds{}
\usepackage{color}

\usepackage{amsmath}
\usepackage{amsthm}
\usepackage{amssymb}
\usepackage{tikz}

\usetikzlibrary{decorations.pathreplacing}
\usepackage{braket}
\usepackage{dsfont}

\usepackage{color}
\usepackage{lineno}
\usepackage{etoolbox}

\usepackage{multirow}

\newcommand{\rom}[1]{\uppercase\expandafter{\romannumeral#1\relax}}

\begin{document}

\preprint{APS/123-QED}

\title{Distributed variational quantum computing with deterministic entanglement tuning}

\author{Ilhwan Kim}
\affiliation{Center for Quantum Technology, Korea Institute of Science and Technology (KIST), Seoul, 02792, Korea}
\affiliation{Department of Applied Physics, Kyung Hee University, Yongin-si, 17104, Korea}

\author{Yong-Su Kim}
\affiliation{Center for Quantum Technology, Korea Institute of Science and Technology (KIST), Seoul, 02792, Korea}
\affiliation{Quantum Information, KIST School, Korea University of Science and Technology, Seoul, 02792, Korea}

\author{Kwang Jo Lee}
\affiliation{Department of Applied Physics, Kyung Hee University, Yongin-si, 17104, Korea}

\author{Hyukjoon Kwon}
\affiliation{School of Computational Sciences, Korea Institute for Advanced Study, Seoul, 02455, Korea}

\author{Yosep Kim}
\thanks{Contact author: \href{mailto:yosep9201@gmail.com}{yosep9201@gmail.com}}
\affiliation{Department of Physics, Korea University, Seoul, 02841, Korea}

\author{Hyang-Tag Lim}
\thanks{Contact author: \href{mailto:hyangtag.lim@kist.re.kr}{hyangtag.lim@kist.re.kr}}
\affiliation{Center for Quantum Technology, Korea Institute of Science and Technology (KIST), Seoul, 02792, Korea}
\affiliation{Quantum Information, KIST School, Korea University of Science and Technology, Seoul, 02792, Korea}

\date{\today}% It is always \today, today,

\begin{abstract}
Distributed quantum computing offers a scalable route to quantum information processing by entangling spatially separated processors. Although gate teleportation enables universal computation across distributed nodes, it requires repeated consumption of high-fidelity Bell pairs, ancillary qubits, and real-time feedforward, which imposes significant overhead and reduces fidelity. However, many variational quantum algorithms do not demand full universality; rather, they rely on sufficient expressibility to explore solution spaces effectively. Building on this, we propose a distributed variational quantum computing protocol based on deterministic entanglement tuning. In contrast to probabilistic filtering, our approach deterministically modulates pre-shared entanglement using only local operations and classical communication. We validate the protocol through a proof-of-principle experiment by estimating ground-state energies of the He-H$^+$ molecule and the Schwinger model, showing that diverse entanglement levels can be engineered to match problem-specific requirements. Our results suggest a practical alternative for near-term distributed quantum applications.
\end{abstract}


\maketitle

\section{Introduction}
Quantum computing promises transformative advances in areas such as chemistry, optimization, and materials science~\cite{Cao2019,Zhou2020,McArdle2020}. However, scaling up quantum systems remains a fundamental challenge due to growing error rates and control complexity in monolithic architectures~\cite{VanMeter2016,Knill1998}. To address this, modular quantum architectures have been proposed, in which multiple quantum processing units (QPUs) are interconnected via quantum channels~\cite{Kimble2008, Monroe2014}. This paradigm gives rise to distributed quantum computing (DQC), where quantum algorithms are executed coherently across spatially separated nodes~\cite{Peng2020, Eddins2022, Avron2021, main2025distributed,hwang2024distributed}.

A key requirement in DQC is the transmission of quantum information between remote nodes, typically accomplished through shared entanglement~\cite{horodecki2009quantum,Chitambar2019}. Quantum gate teleportation, which enables universal nonlocal operations by consuming Bell pairs and applying classical feedforward~\cite{gottesman1999demonstrating,eisert2000}, has been demonstrated across various platforms, including photonic~\cite{liu2024nonlocal,aghaee2025scaling}, trapped-ion~\cite{main2025distributed}, and superconducting systems~\cite{chou2018deterministic}. However, the need to consume an entangled pair for every nonlocal gate imposes substantial resource overhead and accumulates operational errors, limiting scalability in near-term hardware.

Variational quantum algorithms (VQAs) provide an alternative framework, using parameterized quantum circuits that do not require full gate universality~\cite{mcclean2016theory, peruzzo2014variational,Hu2025}. These algorithms depend instead on the expressibility of the chosen ansatz to explore the relevant solution space~\cite{cerezo2021variational}.  This relaxed requirement allows for simpler and more practical distributed implementations~\cite{hwang2024distributed}. A significant obstacle, however, is the limited ability to tune the entanglement between distributed QPUs. Since local unitary operations cannot change the amount of shared entanglement~\cite{nielsen1999conditions}, previous approaches have relied on probabilistic local filtering~\cite{Verstraete2001} or entanglement filtering~\cite{hofmann2002quantum}. Although experimentally demonstrated~\cite{okamoto2009entanglement,Li2022,wang2024_2,Lee2026,Sangouard2011}, such techniques are inherently lossy and non-deterministic, reducing efficiency and scalability.

In this work, we present a protocol for distributed variational quantum computing (DVQC) that enables deterministic entanglement tuning based on Nielsen's majorization criterion~\cite{nielsen1999conditions, Wu2009, Wu2010}. Our method uses only local operations and classical communication (LOCC), and does not rely on postselection. We implement this protocol in a photonic DVQC setup as a proof-of-principle experiment, targeting two benchmark problems with distinct entanglement requirements: the He-H$^+$ molecular model~\cite{kassal2011simulating} and the Schwinger model describing electron-positron interactions~\cite{zohar2012simulating, klco2018quantum}. In both cases, the experimentally measured ground-state energies show close agreement with theoretical predictions across varying degrees of entanglement. These results demonstrate that deterministic entanglement tuning enables efficient preparation of problem-specific variational states in distributed settings, offering a practical and scalable route for near-term DQC.

\begin{figure*}[t!]
	\centering
	\includegraphics[width=0.95\textwidth]{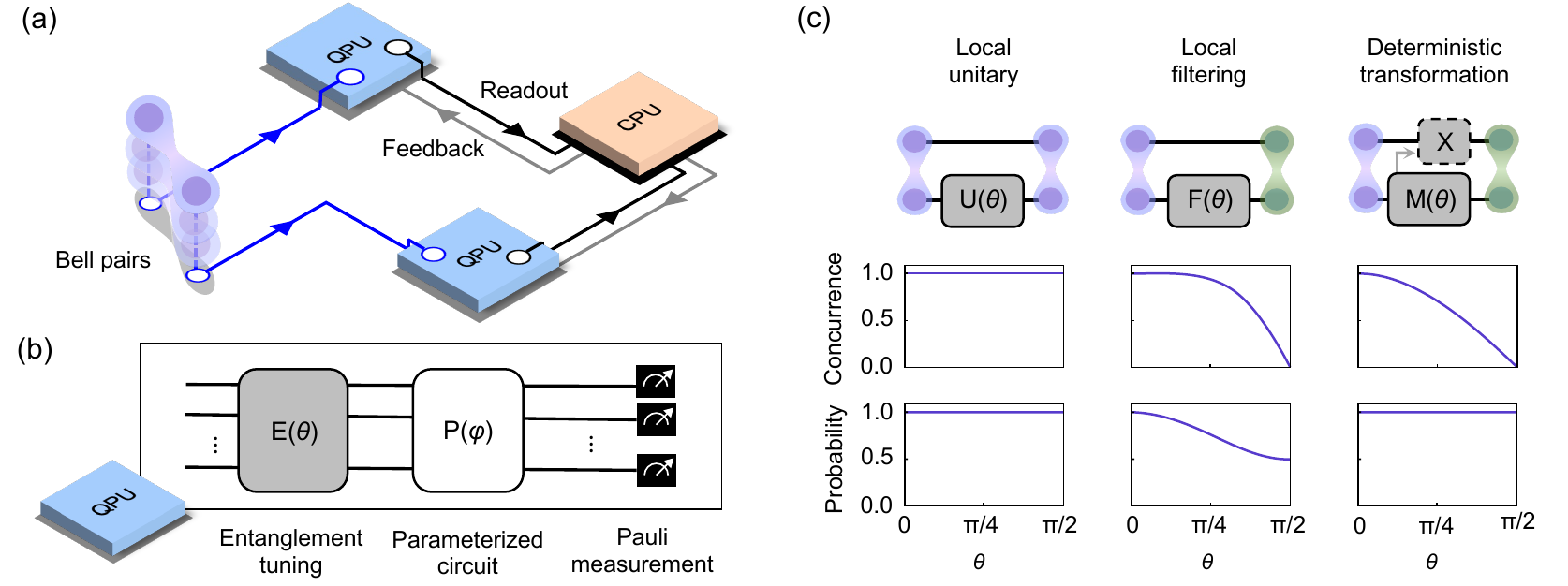}
	\caption{DVQC protocol schematic and entanglement tuning.
		(a) Maximally entangled photon pairs are distributed to two distant QPUs, where local operations and measurements are performed. The readout results are sent to a classical processing unit (CPU), which optimizes circuit parameters and sends updated values back to the QPUs. This feedback loop iteratively refines the quantum state toward the solution.
		(b) Each QPU applies an entanglement tuning operation  $\mathrm{E}(\theta)$, followed by a parameterized quantum circuit $\mathrm{P}(\varphi)$ for state preparation and Pauli measurements. The parameters $\theta$ and $\varphi$ are updated via classical feedback. 
		(c) Concurrence and success probability for three operations applied to the Bell state: a local unitary $\mathrm{U}(\theta)$, a local filtering $\mathrm{F}(\theta)$, and a measurement $\mathrm{M}(\theta)$ followed by a conditional Pauli-X gate. See main text for details.}
	\label{fig1}
\end{figure*}

\section{results}
\subsection{DVQC protocol and entanglement tuning}
Figure~\ref{fig1}(a) illustrates our DVQC protocol. The procedure begins with the distribution of maximally entangled photon pairs, which are then transformed to a target entanglement level via LOCC. This entanglement tunability enables the preparation of variational quantum states across spatially separated QPUs. As shown in Fig.~\ref{fig1}(b), each QPU applies an entanglement tuning operation $ \mathrm{E}(\theta) $, followed by a parameterized ansatz circuit $\mathrm{P}(\varphi)$. The variational parameters $\theta$ and $\varphi$ are iteratively updated using classical feedback from an optimizer running on a CPU.

For clarity, we focus on a bipartite two-qubit example and compare three operations applied to the Bell state $|\psi^+\rangle_{AB} = (|00\rangle + |11\rangle)/\sqrt{2}$ [Fig.~\ref{fig1}(c)]. Extensions to larger systems are discussed later. To quantify the resulting bipartite entanglement, we use concurrence as a standard metric~\cite{wootters1998entanglement}. Local unitary operations preserve entanglement, yielding constant concurrence and unit success probability~\cite{Kraus2010_1,Kraus2010_2}. In contrast, local filtering allows for entanglement tuning through non-unitary amplitude scaling:
\begin{equation}
	\mathrm{F}(\theta) = \begin{pmatrix} \cos\theta & 0 \\ 0 & 1 \end{pmatrix}.
\end{equation}
This method is particularly suitable for photonic systems, as it can be readily implemented by partially filtering photons in a specific basis~\cite{Lim2014,Lee2026}. The unnormalized state after applying $ \mathrm{F}^B(\theta) = I \otimes \mathrm{F}(\theta) $ to the Bell state is:
\begin{equation}
	|\psi^+_\mathrm{F}(\theta)\rangle = \mathrm{F}^B(\theta)|\psi^+\rangle = \frac{1}{\sqrt{2}}(\cos\theta\,|00\rangle + |11\rangle),
\end{equation}
with concurrence and success probability given by:
\begin{eqnarray}
	&&\mathcal{C}_\mathrm{F}(\theta) = \frac{2\cos\theta}{1 + \cos^2\theta}, \\
	&&p_\mathrm{F}(\theta) = \langle \psi^+_\mathrm{F}(\theta) | \psi^+_\mathrm{F}(\theta) \rangle = \frac{1 + \cos^2\theta}{2}.
\end{eqnarray}
This shows that local filtering enables continuous entanglement tuning, albeit at the cost of reduced success probability due to postselection.

\begin{figure*}[t]
	\centering
	\includegraphics[width=0.95\textwidth]{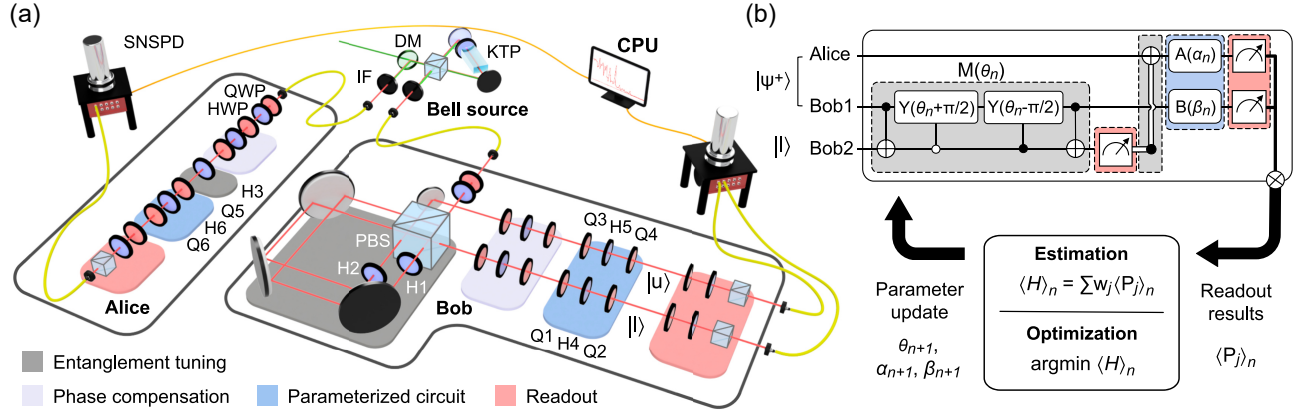}
	\caption{Implementation of DVQC protocol.
		(a) Experimental setup. A polarization-entangled photon pair is generated and distributed to two remote QPUs, labeled Alice and Bob. Bob performs deterministic entanglement tuning using a displaced Sagnac interferometer, which encodes a path-mode ancilla qubit. The positive operator-valued measurement (POVM) $\mathrm{M}(\theta)$ in Eq.~(\ref{eq:measurement}) is implemented by rotating H1 and H2. Depending on whether Bob's photon is detected in path mode $|u\rangle$ or $|l\rangle$, a conditional Pauli-X gate is applied to Alice's photon via H3. After phase compensation, parameterized local circuits are applied on both sides, followed by measurements in the Pauli basis.  Detection outcomes are sent to a CPU for variational parameter updates.
		(b) Quantum circuit diagram. Readout results are used to estimate the target Hamiltonian $\langle H \rangle_n = \sum_j w_j \langle P_j \rangle_n$, and the parameters $\theta_{n+1}, \alpha_{n+1}, \beta_{n+1}$, corresponding to entanglement tuning and the local ansatz, are updated through a classical optimization loop. DM, dichroic mirror; IF, interference filter; HWP, half-wave plate; QWP, quarter-wave plate; PBS, polarizing beam splitter; and SNSPD, superconducting nanowire single-photon detector.}
	\label{fig2}
\end{figure*}

To overcome the limitations of probabilistic entanglement tuning, we introduce a deterministic entanglement transformation protocol based on Nielsen's majorization theorem~\cite{nielsen1999conditions}, which provides necessary and sufficient conditions for converting one pure bipartite entangled state into another using LOCC. The transformation begins with a two-outcome POVM applied locally to one qubit of the maximally entangled Bell state $ |\psi^+\rangle_{AB} $. The Kraus operators corresponding to measurement outcomes 0 and 1 are defined as:
\begin{equation}
	\mathrm{M}_0(\theta) = \begin{pmatrix} \cos\theta' & 0 \\ 0 & \sin\theta' \end{pmatrix}, \ \! 
	\mathrm{M}_1(\theta) = \begin{pmatrix} 0 & \cos\theta' \\ \sin\theta' & 0 \end{pmatrix},\label{eq:measurement}
\end{equation}
where $ \theta' = \theta/2 + \pi/4 $, and the completeness relation $ \sum_i \mathrm{M}_i^\dagger \mathrm{M}_i = I $ is satisfied. 

After the measurement, a classical message is sent to the remote QPU. If the outcome corresponds to $ \mathrm{M}_1 $, a Pauli-X gate is applied to ensure that the resulting state matches the one obtained from $ \mathrm{M}_0 $. The overall transformation is then given by:
\begin{eqnarray}\label{eq:feedforward}
	|\psi^+_\mathrm{M}(\theta)\rangle\langle\psi^+_\mathrm{M}(\theta)| &=& (I \otimes \mathrm{M}_0(\theta)) |\psi^+\rangle\langle\psi^+| (I \otimes \mathrm{M}_0^\dagger(\theta)) \\
	&&+ (X \otimes \mathrm{M}_1(\theta)) |\psi^+\rangle\langle\psi^+| (X \otimes \mathrm{M}_1^\dagger(\theta)). \nonumber
\end{eqnarray}
This process deterministically yields the pure state:
\begin{equation}
	|\psi^+_\mathrm{M}(\theta)\rangle = \cos{\left(\tfrac{\theta}{2}+\tfrac{\pi}{4}\right)}|00\rangle + \sin{\left(\tfrac{\theta}{2}+\tfrac{\pi}{4}\right)}|11\rangle,\label{eq:output_state}
\end{equation}
with concurrence and success probability given by:
\begin{eqnarray}
	&&\mathcal{C}_\mathrm{M}(\theta) = \cos\theta, \\
	&&p_\mathrm{M}(\theta) = 1.
\end{eqnarray}

Unlike local filtering, this transformation enables continuous entanglement tunability while preserving both the input Bell resource and unit success probability. However, Nielsen’s theorem dictates that such transformations are only possible when the Schmidt coefficients of the target state are majorized by those of the initial state, implying that entanglement can only decrease or remain unchanged under LOCC~\cite{nielsen1999conditions}. We therefore suggest starting the DVQC protocol from an entangled resource sufficient to prepare the target state. In particular, sharing $N$ Bell pairs between two $N$-qubit processors provides a bipartite maximally entangled resource, from which the inter-processor entanglement can be arbitrarily tuned via LOCC with linear resource overhead~\cite{Torun2015} (see Supplemental Material~\cite{SM}). Moreover, distributing Bell pairs across multiple nodes enables efficient construction of arbitrary entanglement graph states~\cite{Meignant2019,Chelluri2024}, whose entanglement can be further tuned via LOCC~\cite{Turgut2010,Tajima2013,Torun2019}.

\subsection{Experimental schematic}
We now explain our experimental schematic. Figure~\ref{fig2}(a) shows the linear-optical platform used to implement the DVQC protocol. A polarization-entangled Bell pair $|\psi^+\rangle_{AB} = (|\mathrm{HH}\rangle + |\mathrm{VV}\rangle)/\sqrt{2}$ is generated via the spontaneous parametric down-conversion (SPDC) process in a potassium titanyl phosphate (KTP) crystal embedded in a Sagnac interferometer~\cite{kim2025robust} and distributed to two spatially separated nodes, Alice and Bob. While such photonic states can, in principle, be transferred to solid-state or atomic quantum processors~\cite{Clausen2011,Moehring2007}, all quantum operations in this work are performed directly on photonic qubits, with polarization serving as the encoding basis.

To implement the two-outcome POVM in Eq.~(\ref{eq:measurement}) non-destructively on Bob's qubit, we encode an ancilla qubit in the photon's path degree of freedom, $|u\rangle$ and $|l\rangle$. The quantum circuit in Fig.~\ref{fig2}(b) shows the POVM interaction between Bob's polarization qubit (Bob1) and path qubit (Bob2), which is realized using a displaced Sagnac interferometer~\cite{Lim2014}. Half-wave plates H1 and H2, placed in the counter-propagating paths, are rotated by angles that differ by $\pi/4$; their common angle $\theta$ sets the measurement strength and thus the amount of entanglement~\cite{Wu2009}. The click of one of Bob's two detectors, shown in Fig.~\ref{fig2}(a), determines whether the operation $\mathrm{M}_0(\theta)$ or $\mathrm{M}_1(\theta)$ is applied to the polarization qubit. To complete the deterministic transformation, a conditional Pauli-X operation is applied to Alice's photon when Bob's photon is detected in the path mode $|u\rangle$, as described in Eq.~(\ref{eq:feedforward}). This operation can be implemented by conditionally rotating H3 by $\pi/4$ using a motorized stage on Alice's side. In the present proof-of-principle implementation, real-time feed-forward is not implemented.

For DVQC readout, we configure the measurement basis with half-wave plates, quarter-wave plates, and polarizing beam splitters, then record coincidence counts between Alice's detector and Bob's $|u\rangle$ and $|l\rangle$ path-mode detectors. These counts are summed and used to update the variational parameters $\theta_n$, $\alpha_n$, and $\beta_n$ that control the waveplates H1-H6 and Q1-Q6 at each iteration $n$ of the classical optimization loop. Further experimental details are provided in Appendix A and Appendix B.

\begin{figure}[t]
	\centering
	\includegraphics[width=0.95\columnwidth]{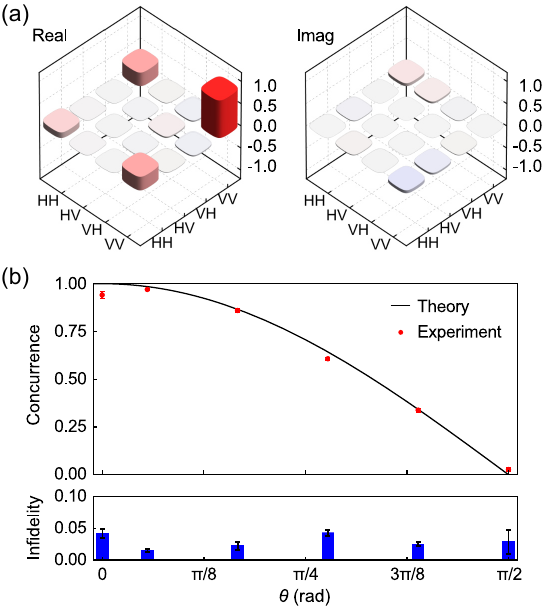}
	\caption{Experimental results of entanglement tuning.
		(a) Experimentally reconstructed density matrix of a partially entangled state prepared by applying the deterministic transformation in Eq.~(\ref{eq:feedforward}) to a Bell state, with $\theta = 5\pi/18$.
		(b) Concurrence and state infidelity of the output states as a function of $\theta$. Red dots represent experimental data, and the black curve indicates the theoretical prediction. Error bars denote statistical uncertainties estimated via Monte Carlo sampling based on Poissonian photon count distributions.
	}
	\label{fig3}
\end{figure}

%%%%%%%%%%%%%%%%%%%%%%%%%%%%%%%%%%%%

\begin{figure}[t]
	\centering
	\includegraphics[width=0.95\columnwidth]{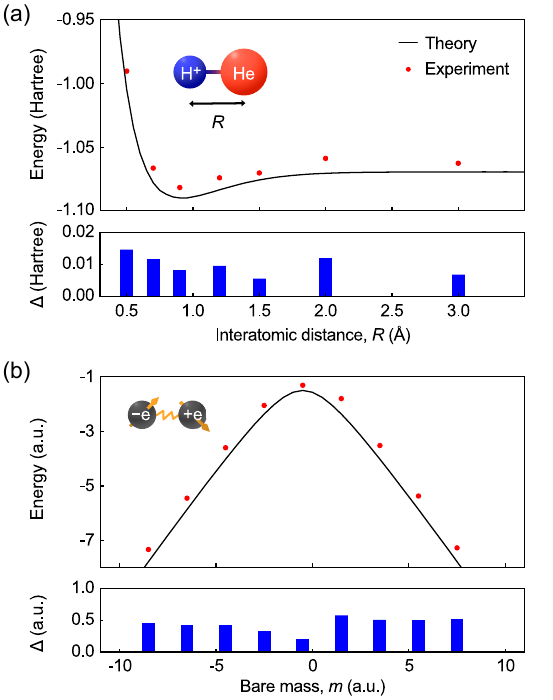}
	\caption{Experimental results of distributed variational quantum eigensolver (DVQE).
		Ground-state energy estimation for (a) the He-H$^+$ molecular system as a function of interatomic distance $R$, and (b) the two-qubit Schwinger model as a function of the bare mass $m$ of the electron and positron. In both cases, the DVQE protocol was implemented with classical optimization over variational parameters, including both local circuit rotations and entanglement tuning. For each value of $R$ and $m$, the experiment was repeated 3 times, typically converging within 50 iterations. The lowest  energy results are shown as red circles, while black curves denote exact theoretical predictions. Lower panels show the energy deviation $\Delta$ between experimental and theoretical values, displayed as blue bars.
	}
	\label{fig4}
\end{figure}

\subsection{Experimental results}
We begin by demonstrating deterministic entanglement tuning starting from the maximally entangled Bell state $|\psi^+\rangle_{AB}$. For the demonstration, all waveplate settings were fixed to identity operations except those used for entanglement tuning and readout. The feed-forward operation on Alice is implemented by correcting her measurement basis offline when Bob’s photon is detected in the path mode $|u\rangle$. The output states were characterized via full quantum state tomography~\cite{James2001}. As a representative example, Fig.~\ref{fig3}(a) presents a partially entangled state corresponding to Eq.~(\ref{eq:output_state}) with $\theta = 5\pi/18$. The resulting concurrence and state infidelity are reported in Fig.~\ref{fig3}(b) as a function of the control parameter $\theta$. The measured concurrence closely follows the theoretical prediction, and the infidelity remains below 0.05 across the entire range, confirming both high-fidelity state preparation and continuous entanglement tunability. Details of the tomographic reconstruction and additional data are provided in Supplemental Material~\cite{SM}.

We then implement a DVQE to estimate ground state energies of two benchmark Hamiltonians: the He-H$^+$ molecular system~\cite{peruzzo2014variational} and the two-qubit Schwinger model~\cite{Borzenkova2021}. The target Hamiltonians are:
\begin{eqnarray}
	&&H_\mathrm{Mol}(R) = \sum_{ij}w_{ij}(R)P^{A}_i P^{B}_j,\label{quantum chemistry hamiltonian} \\
	%\noindent{\rm and,} \nonumber \\
	&&H_\mathrm{Sch}(m) = I^AI^B+X^AX^B+Y^AY^B+\frac{1}{2}Z^AZ^{B} \nonumber \\
	&&\quad \quad \quad \quad \quad-\frac{m+1}{2}Z^AI^B +\frac{m}{2}I^{A}Z^{B},\label{Schwinger Hamiltonian}
\end{eqnarray}
where $P_{i,j} \in \{I, X, Y, Z\}$ are Pauli operators, $R$ denotes the interatomic distance, and $m$ is the bare mass of the electron and positron. Determining ground-state energies provides insight into stable molecular configurations and low-energy quantum field dynamics. Model details and coefficients $w_{ij}(R)$ are given in Supplemental Material~\cite{SM}. 

For a given Hamiltonian, the DVQE algorithm minimizes the expectation value $\langle H \rangle$ by optimizing a parameterized quantum ansatz. The variational ansatz is prepared by applying the entanglement tuning operation $\mathrm{M}(\theta_n)$ followed by local unitaries $A(\alpha_n)$ and $B(\beta_n)$ to the shared Bell state. Optimization is performed via the COBYLA algorithm, which demonstrated reliable convergence with fewer iterations than alternative methods~\cite{Lee2022,Kim2024_2}. The parameter update of $\theta_n$, $\alpha_n$, and $\beta_n$ terminates once the sum of their changes falls below $10^{-6}$.

Figure~\ref{fig4}(a) shows the DVQE results for He-H$^+$, with the lowest ground state energy reaching $E_{g,\text{Mol}} = -1.083\pm 0.001$ Hartree at $R = 0.9$~\AA, in close agreement with the theoretical minimum of $-1.089$ Hartree. The mean deviation over all distances is $\langle \Delta_\mathrm{Mol}\rangle = 0.010 \pm 0.003$ Hartree. For the Schwinger model, Fig.~\ref{fig4}(b) displays the estimated ground-state energies as a function of bare mass $m$. The average error is $\langle \Delta_\mathrm{Sch}\rangle = 0.429 \pm 0.114$, corresponding to a normalized accuracy $\langle\delta_\mathrm{Sch}\rangle = \langle\Delta_\mathrm{Sch}\rangle / (E_1 - E_0) < 0.079 \pm 0.033$, with $E_1 = 1$~\cite{Borzenkova2021}. The deviations may be attributed to state infidelity associated with imperfect Bell-state preparation and residual phase instability between Bob’s two path modes. As the optimal states span a wide range of entanglement, these results confirm that deterministic entanglement tuning provides access to a broad variational space. Exact solution states and their concurrences, along with the corresponding optimization trajectories, are summarized in Supplemental Material~\cite{SM}.

\section{Discussion}

In conclusion, we have presented a scalable and practical approach to DVQC, enabling deterministic and lossless control of entanglement between remote nodes. Unlike gate-teleportation protocols, whose Bell-pair cost scales with circuit depth~\cite{gottesman1999demonstrating,eisert2000}, our method decouples entanglement consumption from circuit depth, making it particularly advantageous in regimes requiring deep variational circuits for higher accuracy~\cite{Bravo-Prieto2020}. As a proof-of-principle demonstration,  we estimated ground-state energies for two representative models--the He-H$^+$ molecule and the two-qubit Schwinger model--spanning a broad range of entanglement regimes. The high concurrence and fidelity observed across these tests validate the precision of our entanglement tuning method, which is critical for preparing accurate, problem-specific quantum states in distributed settings. 

While our current implementation uses two single-qubit photonic processors, the protocol naturally extends to more complex systems. 
The entanglement between two multi-qubit processors can be arbitrarily tuned by pre-sharing a sufficient number of Bell pairs, with a resource overhead that scales linearly with the system size~\cite{Torun2015,SM}. In multipartite systems, entanglement transformation under LOCC becomes more restrictive~\cite{Dur2000,Hebenstreit2022}. Nevertheless, this problem can be detoured by first generating a suitable entanglement structure, such as GHZ, W, or graph states~\cite{Meignant2019,Chelluri2024}, from distributed Bell pairs across multiple quantum processors, and then tuning the amount of entanglement deterministically via LOCC~\cite{Turgut2010,Tajima2013,Torun2019}. Furthermore, the use of pre-shared entanglement in our DVQC protocol permits offline distillation or purification~\cite{Hu2021,Bennett1996,pan2001entanglement}, and correlation-informed qubit permutations that minimize inter-party correlations can further reduce the required Bell-pair budget~\cite{Tkachenko2021}.

Variational quantum algorithms aim to explore structured, physically motivated subspaces rather than arbitrary quantum states; accordingly, restricting the search to LOCC-accessible entangled subspaces enhances scalability and trainability~\cite{Holme2022,Cerezo2022,Lierta2021}. This approach is particularly effective when prior information about the target state’s entanglement amount or structure is available, for example through a known tensor-network representation~\cite{Liu2019,Ho2019}. Taken together, these considerations show that entanglement tuning via LOCC offers a flexible, resource-efficient, and scalable framework for implementing distributed quantum algorithms under near-term experimental constraints.

\section*{Data availability}
The data that support the findings of this article are not publicly available. The data are available from the authors upon reasonable request.

\appendix

\section{Bell source preparation}

We generate the initial Bell state $|\psi^+\rangle_{AB} = (|\mathrm{HH}\rangle + |\mathrm{VV}\rangle)/\sqrt{2}$ using a bidirectional SPDC process in a 20-mm-long bulk KTP crystal embedded in a Sagnac interferometer~\cite{kim2025robust}. The crystal is pumped by a continuous-wave laser at 539.64~nm with an output power of 20~mW, producing polarization-entangled photon pairs at 1079.27~nm. The detected count rate is 502,000$\pm$3160~pairs/s, measured at the output of both arms. The photon pairs are characterized via full quantum state tomography, exhibiting high state quality with a purity of 0.9887$\pm$0.0006, concurrence of 0.9869$\pm$0.0006, and fidelity of 0.9888$\pm$0.0003, with respect to the ideal $|\psi^+\rangle_{AB}$ state. The pump laser at 539.64~nm is obtained through second-harmonic generation in a 40-mm-long periodically poled lithium niobate (PPLN) crystal, with a poling period $\Lambda = 7.1$~$\upmu$m, optimized for quasi-phase matching of a continuous-wave laser at 1079.27~nm.

\section{Deterministic entanglement tuning}

To implement the measurement operation $\mathrm{M}(\theta)$, an ancilla qubit is encoded in the path degree of freedom of Bob’s photon via a displaced Sagnac interferometer. A PBS splits the polarization qubit (Bob1), sending $|\mathrm{H}\rangle$ and $|\mathrm{V}\rangle$ into counter-propagating paths $|l\rangle$ and $|u\rangle$, implementing a CNOT gate as shown in Fig.~\ref{fig2}(b). Each path contains an HWP that performs a controlled-$Y$ rotation, set to $\theta + \pi/2$ and $\theta - \pi/2$, respectively. After recombination at the PBS (second CNOT), Bob’s joint polarization-path state is transformed as:
\begin{eqnarray}
	&&|\mathrm{H}\rangle|l\rangle \rightarrow \cos\left(\tfrac{\theta}{2} + \tfrac{\pi}{4}\right) |\mathrm{H}\rangle|l\rangle + \sin\left(\tfrac{\theta}{2} + \tfrac{\pi}{4}\right) |\mathrm{V}\rangle|u\rangle, \nonumber \\
	&&|\mathrm{V}\rangle|l\rangle \rightarrow \cos\left(\tfrac{\theta}{2} + \tfrac{\pi}{4}\right) |\mathrm{H}\rangle|u\rangle + \sin\left(\tfrac{\theta}{2} + \tfrac{\pi}{4}\right) |\mathrm{V}\rangle|l\rangle. \nonumber
\end{eqnarray}
Applying this to the initial Alice–Bob1–Bob2 state of $\tfrac{1}{\sqrt{2}}(|\mathrm{HH}\rangle + |\mathrm{VV}\rangle)|l\rangle$ yields:
\begin{eqnarray}
	&&\tfrac{1}{\sqrt{2}}\left( \cos\left(\tfrac{\theta}{2} + \tfrac{\pi}{4}\right)|\mathrm{HH}\rangle + \sin\left(\tfrac{\theta}{2} + \tfrac{\pi}{4}\right)|\mathrm{VV}\rangle \right)|l\rangle \nonumber \\
	&&+ \tfrac{1}{\sqrt{2}}\left( \sin\left(\tfrac{\theta}{2} + \tfrac{\pi}{4}\right)|\mathrm{HV}\rangle + \cos\left(\tfrac{\theta}{2} + \tfrac{\pi}{4}\right)|\mathrm{VH}\rangle \right)|u\rangle. \nonumber
\end{eqnarray}

Detection of Bob2 in $|l\rangle$ or $|u\rangle$ occurs with equal probability and corresponds to applying $\mathrm{M}_0(\theta)$ or $\mathrm{M}_1(\theta)$, respectively, as defined in Eq.~(\ref{eq:measurement}). To make the transformation deterministic, a Pauli-X gate is applied to Alice’s photon via an HWP set to $\pi/4$ whenever $|u\rangle$ is detected. This yields a partially entangled state regardless of the detection outcome:
\begin{equation}
	|\psi_\mathrm{M}^+\rangle = \cos\left(\tfrac{\theta}{2} + \tfrac{\pi}{4}\right)|\mathrm{HH}\rangle + \sin\left(\tfrac{\theta}{2} + \tfrac{\pi}{4}\right)|\mathrm{VV}\rangle.
\end{equation}
Relative phase differences in the state, caused by experimental imperfections, are compensated using an HWP and two QWPs oriented at 45$\degree$, as shown in Fig.~\ref{fig2}(a).

\end{document}